\documentclass[12pt,preprint]{aastex}
\def\lax {\ifmmode{_<\atop^{\sim}}\else{${_<\atop^{\sim}}$}\fi}  
\def\gax {\ifmmode{_>\atop^{\sim}}\else{${_>\atop^{\sim}}$}\fi}  
\def\gtorder{\mathrel{\raise.3ex\hbox{$>$}\mkern-14mu
             \lower0.6ex\hbox{$\sim$}}}

\def\cm2{cm$^{-2}$}
\def\s1{s$^{-1}$}

\begin{document}

%Submitted to Frontiers in Astronomy and Space Sciences
\title{
%SDSS~J075: BH Mass of the Secondary and Binary Inclination Estimate
SDSS~J075217.84+193542.2: X-ray weighing of a secondary BH}

\author{Lev Titarchuk\altaffilmark{1} and Elena Seifina\altaffilmark{2}}
%$^{1}$Dipartimento di Fisica, University di Ferrara, Via Saragat 1,  I-44122, Ferrara, Italy \\
%$^{2}$Lomonosov Moscow State University, Sternberg Astronomical Institute, Universitetsky Prospect 13, Moscow, 119992, Russia

\altaffiltext{1}{Dipartimento di Fisica, Universit\`a di Ferrara, Via Saragat 1, I-44100 Ferrara, Italy, email:titarchuk@fe.infn.it; George Mason University Fairfax, VA 22030}
\altaffiltext{2}{Moscow M.V.~Lomonosov State University/Sternberg Astronomical Institute, Universitetsky Prospect 13, Moscow, 119992, Russia; seif@sai.msu.ru}

\begin{abstract}
Precise measurements of  black hole (BHs) masses  are necessary to understand the coevolution of these sources and their host galaxies. Sometimes in the center of a galaxy there is not one, but two BHs. The BH duality of the quasar nucleus SDSS~J075217.84+193542.2 (herein SDSS~J0752) was recently proposed based on the observed strict periodicity of optical emission from the source. We tested this assumption using X-ray observations with Swift/XRT (2008--2010). We fitted the SDSS~J075217 spectrum using  a Comptonization model and discovered soft X-ray variability  in the 0.3--10~keV energy range. We pursued a scenario in which two supermassive BHs at the center of SDSS~J0752 form a pair; and the less massive (secondary) BH periodically crosses/punctures the disk around the more massive (primary) BH. We associate these periodic crossings  with tidal disruptions of the disk and, as a consequence, with an increase in X-rays seen as a flare in SDSS~J0752. During such an X-ray flare event (2008--2010), we discovered a change in the source spectral states and the photon index saturation at the $\Gamma\sim3$ level with mass accretion rate  $\dot M$. For BH mass scaling we used sources: OJ~287, M101 ULX--1 and HLX--1 ESO~243--49,  as a reference ones, and found that M$_{SDSS}=9\times 10^7$ solar masses, assuming $d_{SDSS}= 500$ Mpc. Thus, we obtained a lower limit to a BH mass due the unknown inclination. In addition, we used the virial mass of the secondary BH based on $H_\alpha$-line measurements and we estimated the binary's inclination at SDSS~J0752, $i=80^{\circ}$, using a scaling technique.
\end{abstract}


\section{Introduction}

The duality of objects is a fairly common phenomenon in astrophysics. Most stars in the Universe are binary. Duality is also common among galaxies. Over the past few years, the number of detected binary black holes (BHs) in the centers of galaxies has increased sharply. Although such double objects are no longer uncommon, the manifestation of their binarity remains not fully understood. Analysis of the dual nature of BHs in active galactic nuclei (AGNs) confirms the reality of supermassive BH (SMBH) mergers in galactic nuclei. Scientists have long been concerned about the problem of the origin of ultra massive black holes, because due to the matter accretion onto stellar-mass BHs, they could not form, since this would require time longer than the lifetime of the Universe. 
The discovery of binary AGNs, observations of massive BH merger events and confirmations %that 
of the high resulting masses of  such BHs obtained in recent years are a real breakthrough in our understanding the origin of ultra massive BHs through maasive BH mergers in galactic nuclei.

Binary supermassive black holes are a remarkable by-product of galaxy mergers in the hierarchical universe~\citep{Begelman80}. In the very last stage of their orbital evolution, gravitational wave radiation powers the binary inspiral. Periodic X-ray/optical/radio flares from AGNs during their orbital rotation have been proposed as a powerful tool for studying such binary systems~\citep{Chen20,Liu16}. It should be noted that the very idea of binary systems with SMBHs  was expressed by Boris Komberg back in 1967~\citep{Komberg67}. Subsequently, their hypothesis about the possible duality of nuclei in quasars was brilliantly confirmed by observations (Sect.~\ref{results}). However, the methods for establishing the duality of SMBHs is tenuous at best and is based on four criteria \citep{S23}.%[Transections, astro-ph, 2023]. 

%0000
First of all, the duality of galactic nuclei is is in some cases demonstrated by %established 
 analyzing their observed images (for nearby galaxies), which allows us to resolve the double structure of their central regions (for example, NGC~7727, \cite{Voggel22}; NGC~6240, \cite{Kollatschny20}; Mrk~739, \cite{Koss11}; UGC~4211, \cite{Koss23}). For distant AGNs, spectroscopic studies make it possible to establish duality by the characteristic behavior of spectral line systems (SDSS~J1537+044, \cite{Boroson09}; %Todd A. Boroson and Tod R. Lauer, 20??; 
SDSS~J0927+294, \cite{Eracleous12}). It is clear that the described methods are no longer applicable for very distant sources. In fact, the collision of galaxies (and, accordingly, their nuclei) in the past, which led to the formation of a binary from two SMBHs, but there is in no way to indicate the duality of their new ``nucleus'', except perhaps for the presence of ``extra'' jets in 3C~75 \citep{Molnar17} or the strict periodicity of the source flares in OJ~287 \citep{TSC23,Zhang22}.  In such cases, the duality of the nucleus can sometimes be established by individual exotic features, for example, by the same bending of the jets of each of the SMBHs, by the specific shape of the light curve (for example, from periodicity and  from the double-humped flare %burst 
maximum in the light curve) or as a possible way to connect the results on optical, X-ray and radio data for the same object. As an interesting additional criterion for duality, \citet{TSC23} pointed to the duality of SMBHs in an AGN, established by the following peculiarities. Namely, the discrepancy between the BH mass according to X-ray and optical data, as a possible consequence of the duality of the central AGN in M87 (see \cite{TSC23,TSCO20,S23}). %(Titarchuk et al (2023), Seifina 2023).

Binary BHs in distant quasars are of particular interest and complexity. Because of their remoteness, it is impossible to establish the duality of their nuclei by direct imaging or Doppler shift analysis of spectral lines. Therefore, the main method for establishing their duality is through %remains their 
analysis of the periodicity of radiation from such sources, which may reveal %and  
quasi-periodic oscillations (QPOs) of their emission.

Recently, a QPO with a periodicity of 6.4 years was discovered in the quasar SDSS J075217.84+193542.2 (herein SDSS~J0752) by \cite{Zhang22} at a redshift of 0.117 \citep{Paris18}. The discovery of this QPO made this quasar a reliable candidate for the binary systems containing SMBHs. In fact, the BH duality in the SDSS J0752 center is indicated by the detection of two Gaussian components in the broad H$_{\alpha}$ line, with an expected spatial distance  about 0.02 pc between these two central SMBHs. Their virial masses are about $8.8 \times 10^{7} M_{\odot}$ and $1.04 \times 10^{9} M_{\odot}$ (Table~\ref{tab:parameters_sdss}).

This periodicity is derived from analysis of the 13.6-year optical light curve from the Catalina Sky Survey (CSS) \citep{Drake09} and from the All-Sky Automated Survey for Supernovae (ASAS-SN) \citep{Shappee14,Kochanek17}. The 6.4-year QPOs are also confirmed using the generalized Lomb-Scargle periodogram with a confidence level above 99.99\%, as well as using the results of autocorrelation analysis and using the weighted wavelet z-transformation technique.

\begin{figure}
\centering
\includegraphics[width=9cm]{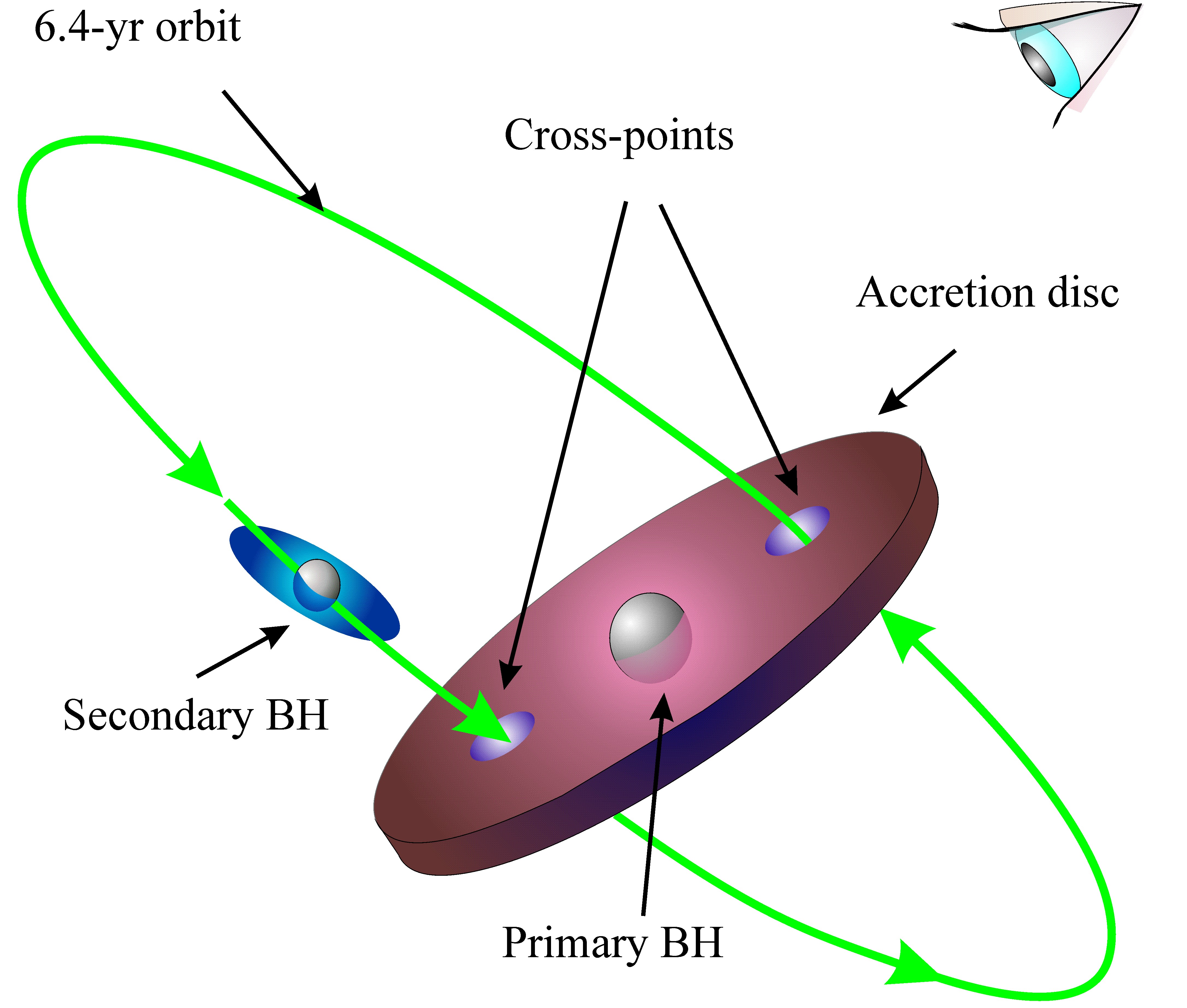}
\caption{Schematic view of SDSS~J0752 model used in our analysis.
}
\label{picture}
\end{figure}

\begin{figure}
\centering
\includegraphics[scale=0.90,angle=0]{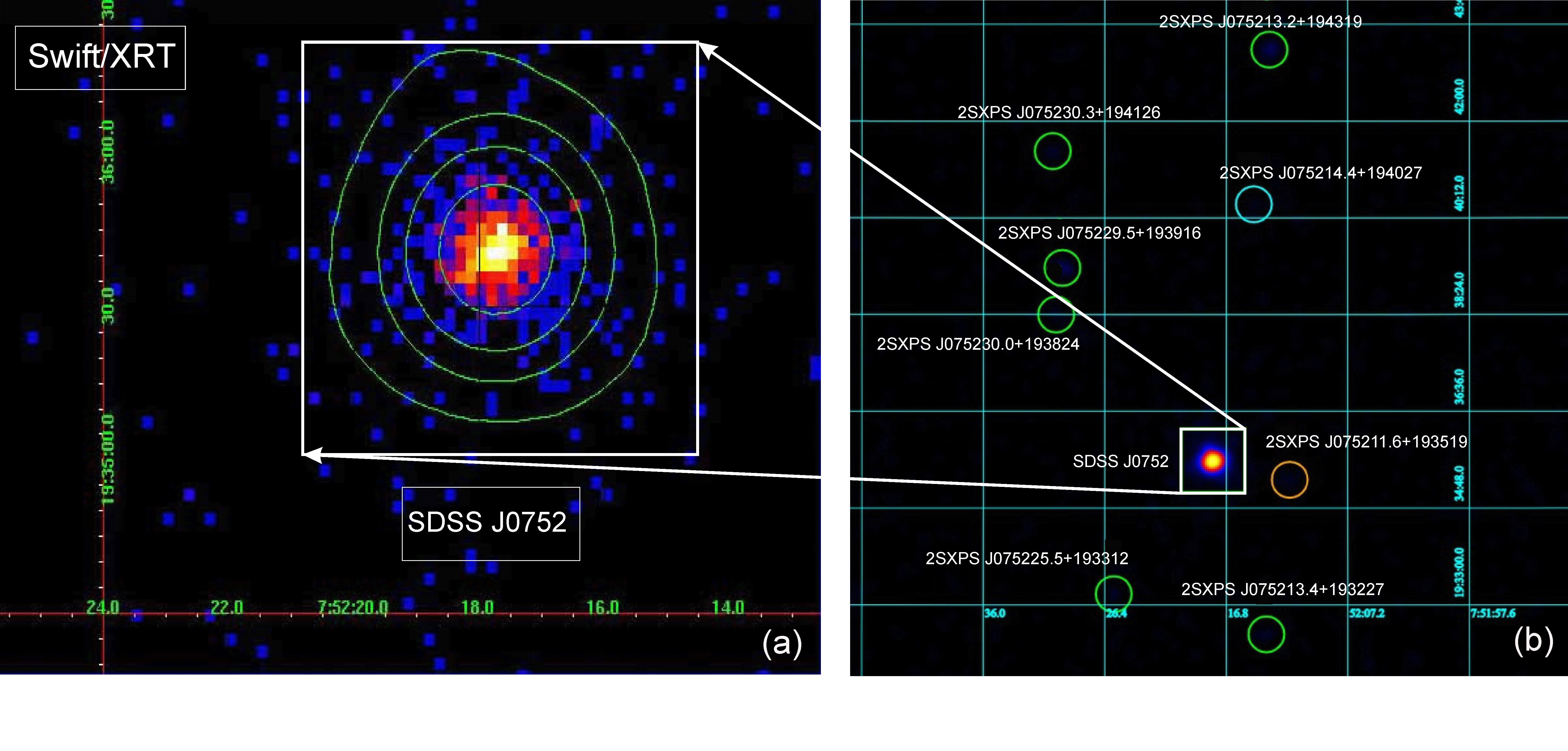}
\caption{$Swift$ X-ray image of SDSS~J0752, accumulated from 2 October 31, 2008 to May 30, 2010 with an exposure time 14 ks. %with a 14 ks exposure. 
Green contours in an enlarged image (a) of SDSS~J0752  confirm the absence of other nearby objects in the 1.3{\tt'} %field of view (
FOV; this image is 160 pixels (=6.3{\tt'} to a side). SDSS %1SXPS 
J075217.7+193540 is indicated by the white box in the panel (b),  while the rest sources of the FOV are marked with circles and corresponding names from the 2SXPS catalogue. The next nearest source (indicated by orange circle) is 86{\tt"} away.
%The contours correspond to fourteen logarithmic intervals with respect to the brightest pixel.
}
\label{imagea}
\end{figure}

In this paper, we test the binary black hole hypothesis in the  SDSS~J0752  (schematically presented in Fig.~\ref{picture}) using Swift/XRT data from an X-ray counterpart of this source (Fig.~\ref{imagea}). The specific goal of our paper is to estimate the mass of the secondary BH in the SDSS~J0752 applying the scaling method to determine the mass of the BH (\cite{st09}, hereafter ST09) based on {\it Swift} observations during X-ray flaring events (Fig.~\ref{Swift_lc}). Based on %Due to 
the discovered more or less strict  6.4-year cycle, we adhere to the hypothesis of orbital rotation of components in a binary black hole, in which the secondary BH periodically disturbs the accretion disk around the primary BH (Fig.~\ref{picture}) to explain the optical (and possibly X-ray) periodic variability of SDSS~J0752 \citep{Zhang22}. In this case, the secondary BH is surrounded by a smaller disk, a minidisk, consisting of matter captured during such periodic passages of the primary BH disk (see Fig.~\ref{picture}).

The scaling method was proposed back in 2007 by \cite{st07}, hereafter ST07, and by ST09. It is worth noting that  there are two   scaling methods:  based on the correlation between the photon index, $\Gamma$, and  the QPO frequency,   $\nu_L$; and one  based on the correlation between $\Gamma$ and the  normalization of the spectrum proportional to $\dot M$. %If, f
For the first {method} ($\Gamma-\nu_L$), the source distance is not required to estimate the black hole mass (ST07), while for the second {method},  $\Gamma-\dot M$  (see ST09), the source distance and  the { inclination}  of the accretion disk relative to the Earth observer  are needed.

 For both methods, it is necessary for the source to show a change in spectral states,  accompanied by a characteristic behavior of the index $\Gamma$, during the flare. %outburst. %and a characteristic behavior of $\Gamma$. \LEt{ This is an incomplete sentence, please rephrase.}An increase of 
 $\Gamma$ monotonically increase  with $\nu_L$ {or} $\dot M$ during the transition from the low hard state (LHS) through the intermediate state (IS) to the high soft state (HSS)
%in the  LHS$\to$IS$\to$HSS transition 
and reaching a constant level  ({saturating}) at high values of  
$\nu_L$ or $\dot M$. Then $\Gamma$ monotonically decreases during a HSS$\to$IS$\to$LHS transition when the flare %ou 
decays. The saturation of $\Gamma$ (so-called $\Gamma$-saturation phase whereby $\Gamma$ reaches a constant value in our parameter space; see Fig.~\ref{three_scal}) 
%***you've not explained this yet, so do so here***) 
during  a flare % outburst 
is a specific signature  that this particular object contains a BH~\citep{tz98}. Indeed,  the $\Gamma$-saturation phase can be only  caused by an accretion flow converging to the event horizon of a BH (see numerical %the Monte-Carlo 
simulation results by  \cite{LT99,LT11}). Then, it makes sense to compare BH sources that have the same $\Gamma$-saturation levels.  In the second method ($\Gamma-\dot M$), it is assumed that  BH luminosity is directly proportional to $\dot M$ (and, consequently, to the mass of the central BH), and inversely proportional to the squared distance to the source. Thus,  we can determine the BH mass by comparisons of  the corresponding tracks    $\Gamma-\dot M$ for a pair of sources with BHs, 
in which 
%parameters are known for one  source, and
 %for the other  one
  all parameters are known except for the BH mass (for more details on the scaling method, see \cite{tss10,ST10} and ST09). 

The scaling approach, in general, has a number of advantages over other methods for determination of a BH  mass. A determination  of the X-ray spectrum arising in the innermost part of the source, based on the  fundamental physical models, taking into account the Comptonization of soft disk photons,
% originated in the  innermost disk part,  
by hot electrons from the Compton cloud  and in the converging flow into a BH. %In fact, in the case of a $10^8 M_{\odot}$ BH, the peak temperature of the disk is relatively low, $ kT_s<1~{\rm keV}/(M_{BH}/10M_{\odot}) ^{1/4}$ , that is, about 20--100 eV, and its thermal peak is in the UV energy range (for details of our observations, see% \S 3 and \cite{ss73}). 
In this case, the latest achievements in modeling the BH states using the Monte-Carlo method were used, based on the numerical solution of the complete relativistic kinetic equation and the detection of the ``saturation'' phase of the spectral index at the maximum of the X-ray flare %outburst 
of the black hole. The method was developed and tested by us on various astrophysical objects and showed excellent agreement with classical methods see e.g. 
\cite{ST10,STS14,TS16,TS17,STV17,SCT18,TSCO20,TS23}. %. The results were published in a series of articles [Seifina and Titarchuk (2010, ApJ, 722, 586); Seifina et al. (2014, ApJ, 789, 57); Titarchuk and Seifina (2016, A&A, 585, 94; 595, 101); Titarchuk and Seifina (2017, A&A, 602, 113); Seifina et al. (2017, A&A, 607, 38); Seifina et al. (2018, A&A, 613, 48; 619, 21); Titarchuk et al. (2020, A&A, 633, 73); Titarchuk and Seifina (2023, A&A, 669, 57; 671, 159). 
The method is based on first principles and fundamental assessments of the gravitational effect of a black hole on the matter surrounding it, as well as on calculating the effective size and mass of the reaction zone to such an effect.
%888888

In this paper, based on  a particular {\it Swift/XRT} data analysis, we estimate a BH mass in SDSS~J0752 using the scaling method. In \S 2 we provide details of our data analysis, while in \S  3 
%\ref{sp_analysis} 
we present a description of the spectral models used for fitting these data. In \S 4 we focus on  the interpretation of our observations. In \S 5 
we focus on the construction  of the power density spectra (PDS) and its interpretation. 
In  \S  6  we  discuss  the  main results of the paper. In \S 7  %\ref{conclusions} 
we present our final conclusions.

\section{DATA REDUCTION \label{data}}

 Using {\it Swift}/XRT data in the 0.3--10 keV energy range, we studied flaring events of SDSS~J0752 from 2008 to 2010 (see the log of observations 
in Table~\ref{tab:list_Swift}).  The data used in this paper are public and available through the GSFC public archive\footnote{https://heasarc.gsfc.nasa.gov}. %In Table~\ref{tab:list_Swift}  we report the log of observations for SDSS~J0752 used in our study.  
%We must admit that not all of the flare events may be related to the secondary BH -- disk  interactions.  Thus, it may be challenging to disentangle them from other flaring events going on in OJ 287 all the time, which may be %is completely unrelated to any binary's presence.
It must be acknowledged  %accepted 
that not all flare events can be associated with the disk--secondary BH interactions. %Therefore, it is difficult to separate 
%them from other 
%flares, that constantly occur in SDSS~J0752, and  completely unrelated to the presence of any binary.
Therefore, it is difficult to separate the sporadic flares that often occur in SDSS~J0752, which are completely unrelated to the presence of any binary  in the SDSS~J0752 center. However, the identified periodicity of the optical source facilitates such identification, since it can be associated mainly with orbital variability. Here we assume that X-ray variability caused by flares due to the passage of a smaller BH through the disk around a larger BH. Moreover, it will be accompanied by corresponding spectral changes, which will help us separate the X-ray flare activity of the small BH. In Sect.~\ref{results} we present the X-ray variability analysis of SDSS~J0752 (using Swift/XRT)  and show that SDSS~J0752 follows spectral patterns typical of galactic and extragalactic sources with BHs. %, which is why we are inclined to identify the detected flare events as caused by the influence of a small BH.

Data were processed using the HEASOFT v6.14, the tool {\tt xrtpipeline} v0.12.84, and the calibration files (CALDB version 4.1). The ancillary response files were created using {\tt xrtmkarf} v0.6.0 and exposure maps were generated by {\tt xrtexpomap} v0.2.7. Source events were accumulated within a circular region with a radius of 50{\tt"} centered at the position of SDSS~J0752 ($\alpha=07^{h}52^{m}17^s.63$ and $\delta=+19^{\circ} 35{\tt '} 42{\tt''}.7$, J2000.0). 
Given the low count rate of SDSS~J0752 %our quasar 
most of the data were collected using the most sensitive Photon Counting (PC) mode.  
%We used XRT data both in %the Windowed Timing %(WT) mode ($\ge$ 1 count/s) and in the Photon Counting %(PC) mode. % for the remaining observations when the X-ray source became sufficiently  faint. 
The background was estimated in a nearby source-free circular region with a 85{\tt"} radius. 

Using  the {\tt xselect} v2.4 task, source and background light curves and spectra were generated. 
Spectra were rebinned with at least ten %\LEt{ Write out numerals when lower than eleven and not directly used as a measurement with the unit following. See Sect 2.7 of the language guide https://www.aanda.org/for-authors/language-editing/2-main-guidelines.}
 counts  in each energy bin using the {\tt grppha} task in order to apply $\chi^2$ statistics. We also used the online XRT data product generator\footnote{http://www.swift.ac.uk/user\_objects/} to obtain the image of the source field of view in order to make a visual inspection and to get rid of possible contamination from nearby sources  \citep{Evans07,Evans09}. The {\it Swift}/XRT (0.3 -- 10 keV) image of the SDSS~J0752 field of view (FOV) is presented in the right panel of Fig.~\ref{imagea}, where %. While 
the panel (a) %of this figure 
demonstrates absence of the X-ray jet-like (elongated) structure as well as the minimal contamination by other point sources and diffuse emission within a region with a 85{\tt"} radius around SDSS~J0752. The next nearest source 2SXPS~J075211.6+193519 is 86{\tt"} away (marked by orange circle). We used $Swift$ observation of SDSS~J0752 (2008 -- 2010) extracted from the HEASARC archives and found that these data  cover a wide range of X-ray luminosities.  

\begin{figure}
\centering
\includegraphics[width=12cm]{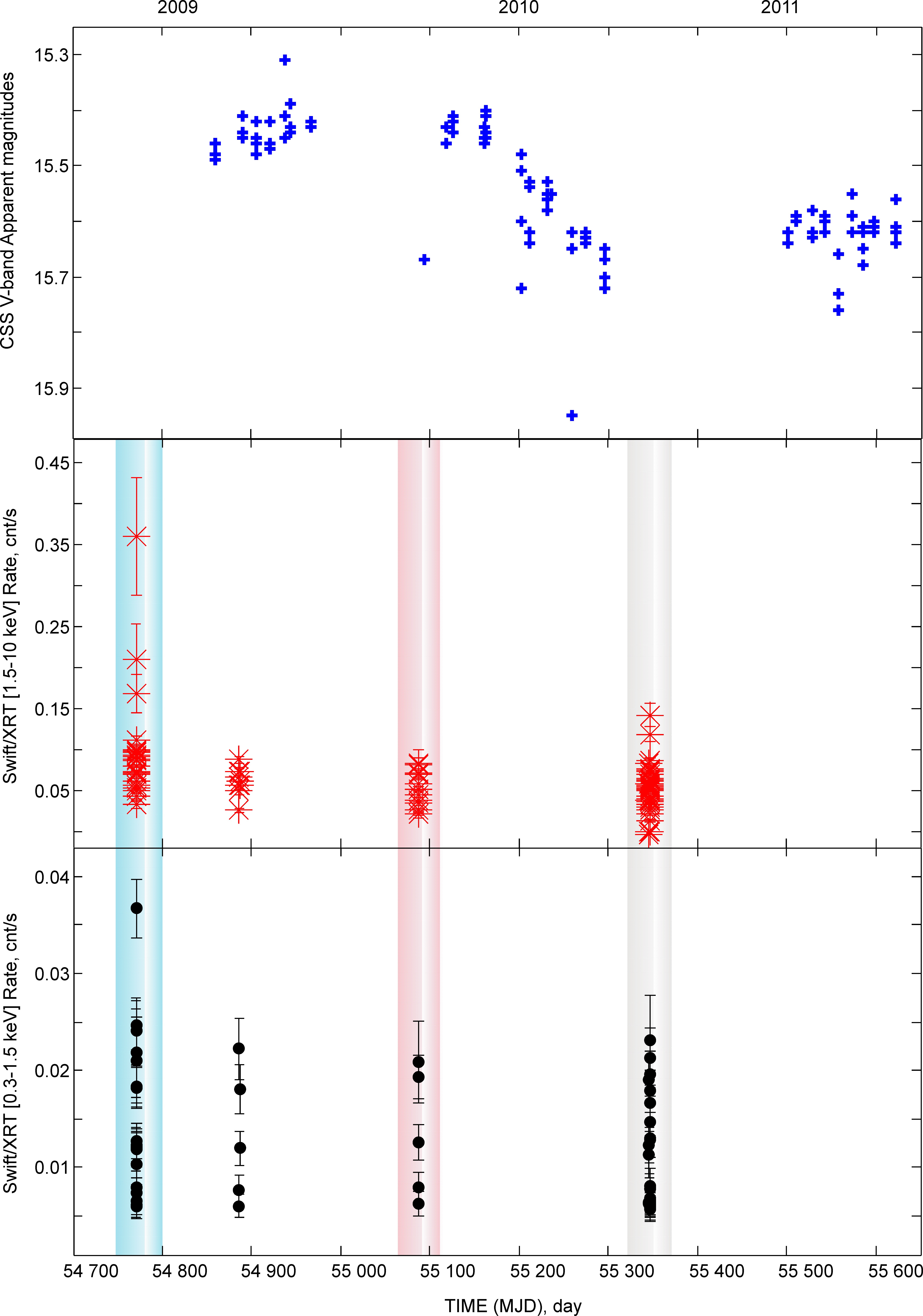}
\caption{Evolution of 
the CSS V-band light curve (upper panel), the XRT/$Swift$ [1.5--10 keV] count rate (middle panel) and the XRT/$Swift$ [0.3--1.5 keV] count rate 
%hardness ratio HR (1.5--10keV/0.3--1.5keV, 
(bottom panel) of SDSS~J0752 from 2008 to 2010. 
The phases LHS, IS and HSS are marked with blue, grey and red vertical strips, for which the corresponding spectra are presented in Fig.~\ref{3_spectra}.
}
\label{Swift_lc}
\end{figure}

Before proceeding and providing the details of the spectral fitting, we study a long-term behavior of SDSS~J0752, 
%OJ~287, 
in particular, its activity patterns. We present a long-term X-ray light curve of  SDSS~J0752 
%OJ~287 
detected by the XRT on board {\it Swift} from 2008 -- 2010 (see Fig.~\ref{Swift_lc}). 

We note that %It is important to note\LEt{ A\&&A discourages authors from directly addressing the reader - for example "note that" can be either deleted completely or replaced with "we note that”.} that 
this X-ray light curve makes it rather difficult to judge the 6.4-yr periodicity found earlier from optical observations (see also the top panel of Fig.~\ref{Swift_lc}). But it can be unequivocally stated that the quasar SDSS~J0752 has become active over the past 15 years and shows sporadic X-ray activity (e.g., MJD 54700--55400, see Fig.~\ref{Swift_lc}).

\section{RESULTS\label{results}}

\subsection{Image of SDSS~J0752\label{image}}
We detected an X-ray source (within the error radius $3.6{\tt ''}$ with 90\% confidence) at the location of SDSS~J0752 indicated by optical observations $RA=118.0735^{\circ}$, $Dec=19.5952^{\circ}$ (refs). At the same time, the coordinates of our X-ray source associated with SDSS~J0752 were some refined $RA=118.0779^{\circ}$ ($\alpha=07^{h}52^{m}18^s.70$), $Dec=+19.6351^{\circ}$ ($\delta=+19^{\circ} 38{\tt '} 06{\tt ''}$, J2000.0) in accordance with the analysis of XTR/Swift data. Visual inspection of the source with the indicated coordinates showed the presence of an almost point-like source in the SDSS~J0752 position %error-radius 
and the absence of other nearby objects within a radius of 85${\tt ''}$.

\subsection{X-ray light curve of SDSS~J0752\label{lc}}
We discovered variability of SDSS~J0752 in %soft 
X-rays. The  Swift/XRT light curves of this source in the 1.5--10 keV energy range (central panel) and in the 0.3--1.5 keV energy range (bottoml panel) from 2008 to 2010 is presented in Fig.~\ref{Swift_lc}. % (central and bottom panels). 
%Here the hardness ratio is defined as (H--S)/(H+S) where H is the 1.5--10 keV count rate and S is the 0.3--1.5 keV count rate.
The figure shows light curves obtained with the Binning mode by time. To test source variability within one observation, we set the time interval size to $\sim$100 s. % or 200 s.
From the figure we can see that the source is variable within one observation, even taking into account observational errors. 
% The figure shows that the source is variable within one observation, even taking into account %measurement (observational errors. 
Fig.~\ref{Swift_lc} also demonstrates the change in the maximum countrate achieved in each observation %(i.e. spacecraft orbit) 
during the full source observation interval 2008--2010. % (Table 1). 
A comparison of the source light curves %of the source 
in different energy bands showed that the main variability of SDSS~J0752 is due to changes in the hard band (1.5--10 keV), although minor variability in the soft band (0.3--1.5 keV) also occurs. 
We also present CSS V-band light curve data, taken from http://nesssi.cacr.caltech.edu/DataRelease/, to compare the variability of SDSS~J0752 in different energy bands for the time interval MJD 54770 (October 2008) to 55650 (March 2011). 
It is clear from the figure that the source is variable in both the X-ray and optical bands. Both bands show the tendency of the source to flare at MJD 54900--55100. In this case, the contribution of hard photons is significant for all states, but for MJD 55100 it has become significantly smaller, which corresponds to a typical flare state for BHs %sources with black holes 
\citep{TSC23} and with a softened X-ray spectrum of the source (for example, see the red spectrum in Fig.~\ref{3_spectra}).
%123
 The phases LHS, IS and HSS are marked with blue, grey and red vertical strips, for which the corresponding spectra are presented in Fig.~\ref{3_spectra} in the next Section.

\subsection{Hardness-intensity diagram of SDSS~J0752\label{HID_1}}
\begin{figure}
\centering
\includegraphics[width=10cm]{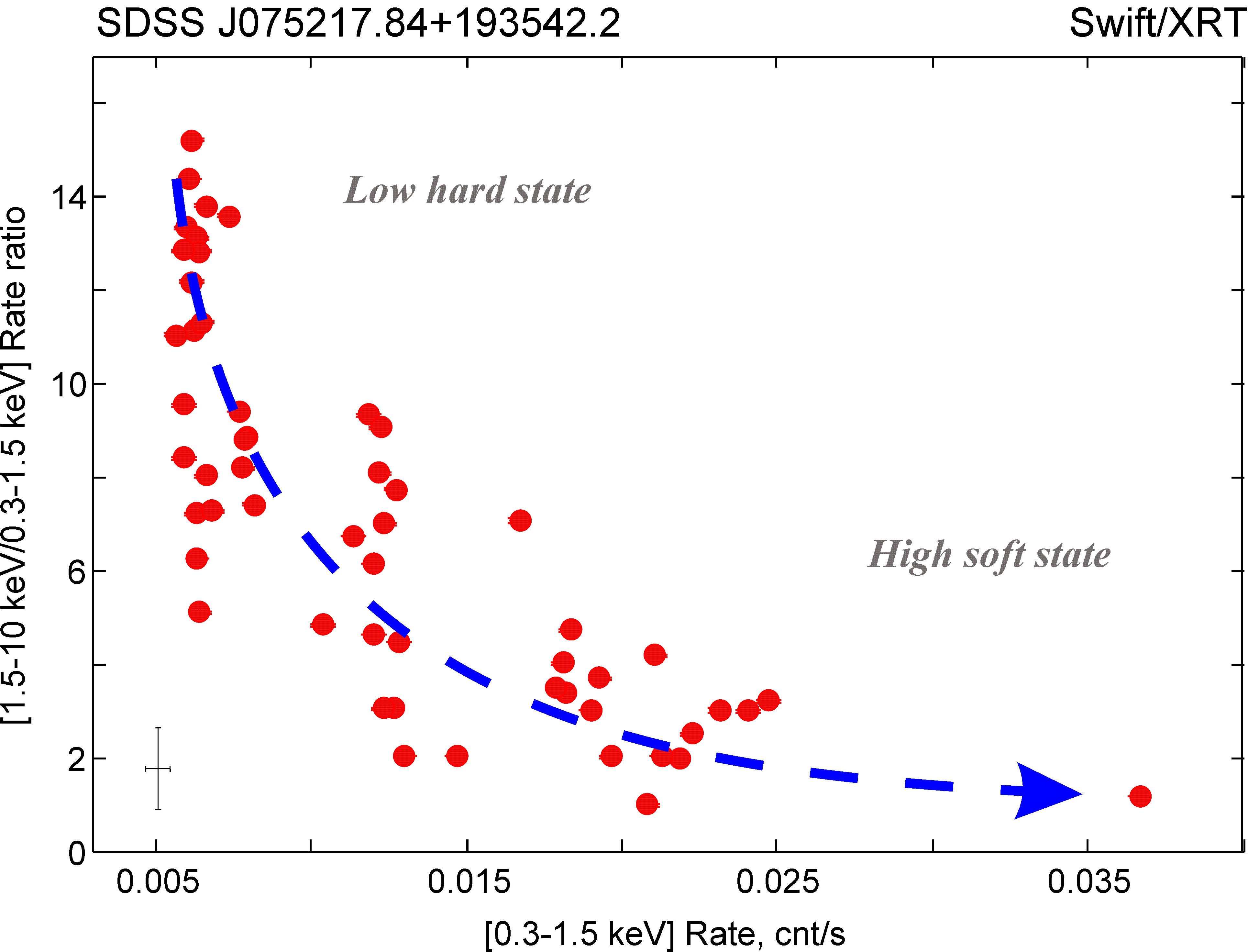}
\caption{
Hardness-intensity diagram (HID) for SDSS~J0752 using the Swift/XRT observations (2008--2009) during spectral evolution from the high state to the low state. In the vertical axis, the hardness ratio (HR) is a ratio of the source counts in the two energy bands: the hard (1.5--10 keV) and soft (0.3--1.5 keV). HR decreases with a soft source brightness in the 0.3--1.5 keV range (horizontal axis). For clarity, we plot only one point with error bars (in the bottom left corner) to demonstrate typical uncertainties for the count rate and HR. The blue dashed arrow indicates the direction of softening of radiation during the transition from LHS to HSS.}
\label{HID}
\end{figure}

\begin{figure}
 \centering
\includegraphics[width=9cm]{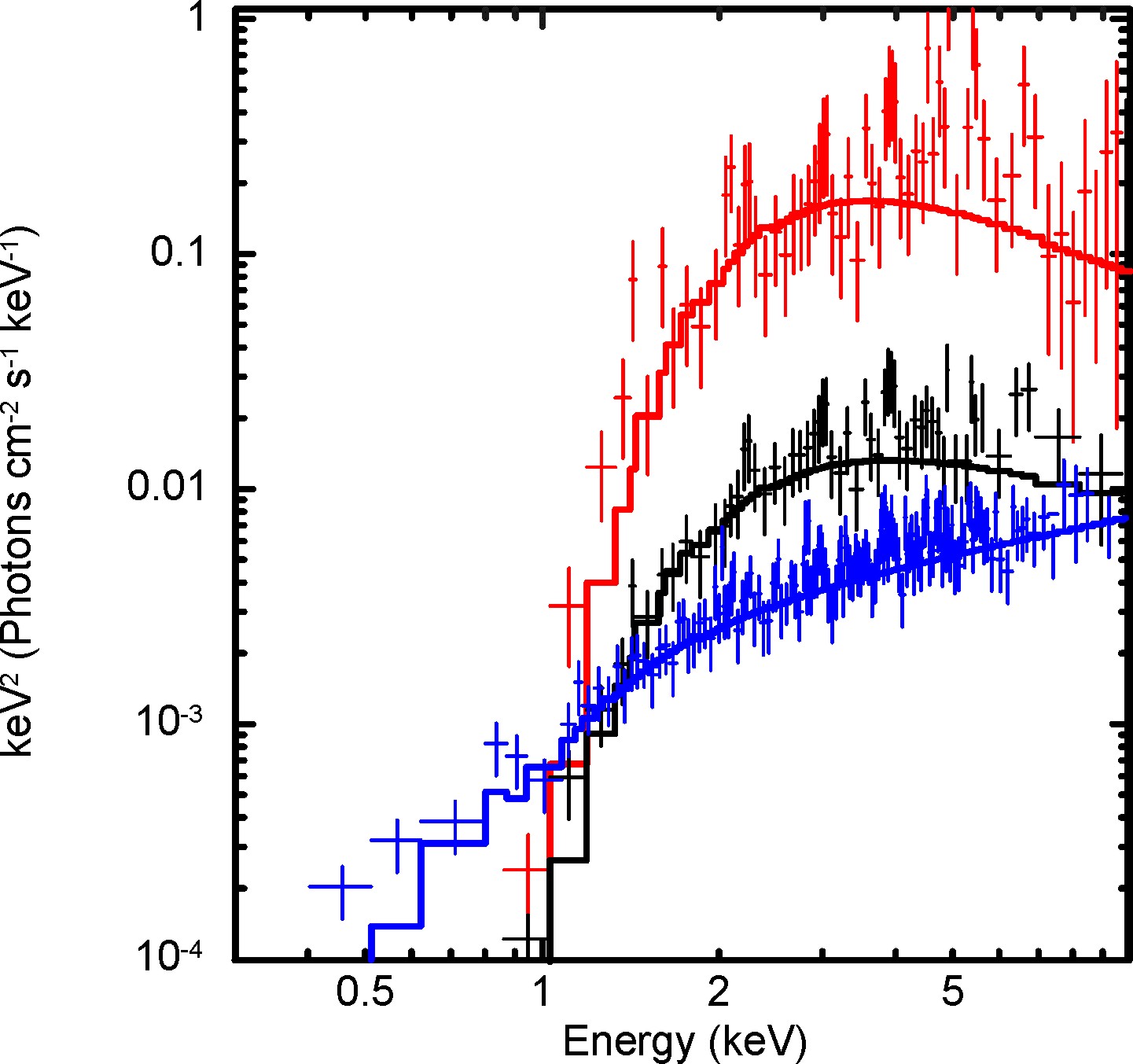}
   \caption{
Three representative spectra of SDSS~J0752 from {\it Swift} data with the best-fit modeling for the LHS (ID=00038041001), IS (ID=00039551001)  and HSS (ID=00038041002) states in units of $E*F(E)$ using the  {\tt tbabs*bmc} model.  The data are denoted by  crosses, while the spectral model is shown by a blue, black and red histograms for each state.
}
\label{3_spectra}
\end{figure}

In application to the Swift data of SDSS~J0752 we have defined the hardness ratio (HR) as a ratio of the hard and soft counts in the 1.5--10 keV and 0.3--1.5 keV bands, respectively. In Figure~\ref{HID} we present hardness-intensity diagram (HID) for SDSS~J0752 using the Swift/XRT observations (2008--2009) during spectral evolution from the high state to the low state. The diagram demonstrates that different count-rate observations are related to different color regimes. The HR larger values correspond to harder spectra. %A Bayesian approach was used to estimate the HR values and their errors (see Park et al. 2006)5. 
For clarity, we plot only one point with error bars (in the bottom left corner) to demonstrate typical uncertainties for the soft count rate and HR. Figure~\ref{HID} clearly shows that the HR monotonically drops with the soft count rate (0.3--1.5 keV). This HID indicates %the 
 X-ray variability of SDSS~J0752 with changing spectral states. This particular sample is similar to those of most of flares %outbursts 
of Galactic X-ray binary transients (see \citet{Homan01,Belloni06,st09,ts09,Shrader10,Munoz-Darias14}).

Below, %Therefore, 
we analyzed the SDSS~J0752 spectrum in each observation to better understand the nature of  this X-ray variability.

\subsection{Spectral Analysis of SDSS~J0752\label{spectral analysis}}

To fit the energy spectra of SDSS~J0752, we used an {\tt XSPEC}   Bulk Motion Comptonization model %consisting of  the  
(hereafter  BMC) %component 
[see \cite{tz98} and \cite{ LT99}]. We also used a multiplicative {\tt tbabs} model \citep{W00} which takes  absorption by neutral material into account. We assume that  accretion onto a BH is described by two main zones (see, for example, Fig.~1 in \cite{TS21}): a geometrically thin accretion disk (e.g., the standard Shakura-Sunyaev disk, see SS73) and a transition layer (TL), which is an intermediate zone between the accretion disk, and a converging (bulk) region  which is assumed  to exist below 3 Schwarzschild radii, $3R_S = 6GM_{\rm BH}/c^2$, [see  details in \cite{tf04}]. The spectral model parameters are  equivalent hydrogen absorption column density $N_H$; the photon index $\Gamma$;  $\log (A),$ which is related to the Comptonized factor $f$ [$={A}/{(1+A)}$]; the  color temperature and normalization of the seed photon blackbody component $kT_s$ and   $N$, respectively.  The parameter $\log(A)$ of the BMC component is fixed at two when the best-fit $\log(A)\gg 1$. In fact, for a sufficiently high $\log(A)\gg1$ (and, therefore, a high value  for $A$), the illumination factor $f$ % = A/(1 + A)$ 
becomes a constant value close to one (that is, the same as in the case of $\log(A) = 2$).  We note that $N_H$ was fixed at the Galactic absorption level of {$2.13\times 10^{22}$ cm$^{-2}$~\citep{W00}. 

Similarly to the  {\it bbody} XSPEC model, the  normalization parameter of the BMC model} is a ratio of the source (disk) luminosity $L$ to the square of the distance $d$  (ST09, see  Eq.~1 there): 

\begin{equation}
N_{bmc}=\biggl(\frac{L}{10^{39}\mathrm{erg/s}}\biggr)\biggl(\frac{10\,\mathrm{kpc}}{d}\biggr)^2.
\label{bmc_norm}
\end{equation}  

This encompasses  an important property of our model. Namely, using this model, one can correctly
evaluate  normalization of the original ``seed'' component, which is presumably a correct $\dot M$ indicator~\citep{ST11}. In turn, %\LEt{ Or "During its turn" depending on the meaning.} 

\begin{equation}
L = \frac{GM_{BH}\dot M}{R_{*}}=\eta(r_{*})\dot m L_{Edd}.
\label{bmc_norm_lum}
\end{equation}  
Here $R_{*} = r_{*} R_S$ is an effective radius where the main energy release takes place in the disk, $R_S = 2GM/c^2$ is the Schwarzschild radius, $\eta = 1/(2r_{*})$, $\dot m = \dot M/\dot M_{crit}$ is the dimensionless $\dot M$ in units of the critical mass accretion rate $\dot M_{crit} = L_{Edd}/c^2$, and $L_{Edd}$ is the Eddington luminosity. For the formulation of the  Comptonization problem,  one can refer to %\LEt{ Idem. as note 23.} 
 \cite{tmk97,tz98,LT99,Borozdin99}, and \cite{st09}.  

Spectral analysis of the $Swift$/XRT data  fits for SDSS~J0752 provides a general picture of the source evolution.  We can trace the change in the spectrum shape during the LHS--IS--HSS transition. In  Fig.~\ref{3_spectra}, we  show  three representative $E*F_E$ spectral diagrams %(red lines) 
for different states of SDSS~J0752. Here, we put together spectra of the LHS, IS, and HSS, to demonstrate the source spectral evolution from the low-hard to   high-soft states to  states based on the $Swift$ observations. The data are presented here as follows: the LHS (taken from observation 00038041001, blue), the IS (00039551001, black) and the HSS (00038041002, red) in units $E*F(E)$ fitted using the  {\tt tbabs*bmc} model.  Exposure times  are  3.6, 1.3 anf 5.0 ks,  respectively. 
It is worth noting, that the source spectra were constructed based on the integral flux during the entire observation %(ID=hh67, ..) 
without taking into account binning to achieve a good signal-to-noise value. These  spectra (blue,  red and black) are plotted for the LHS, IS and HSS phases, marked with blue, gray and red vertical stripes, respectively, in Fig.~\ref{Swift_lc}. 

The best-fit parameters in the HSS state (red spectrum in Fig.~\ref{3_spectra}) are $\Gamma$ = 2.9$\pm$0.6, $kT_s$ = 130$\pm$5 eV, $N$ = 8.2$\pm$0.6 L$_{34}$/d$^2_{10}$, and $\log(A)$ = -1.47$\pm$0.08   %\LEt{ Please check the meaning hasn't changed.}
for which the reduced chi-square value $\chi^2_{red}$=1.04 for 964 degrees of freedom (dof), while the best-fit model parameters for the IS state (black spectrum) are 
$\Gamma$ = 2.2$\pm$0.6, $kT_s$ = 140$\pm$3 eV, $N$ = 2.3$\pm$0.8 L$_{33}$/d$^2_{10}$, and $\log(A)$ = -0.32$\pm$0.09  ($\chi^2_{red}$=0.95 for 288 dof); and, finally, the
best-fit model parameters for the LHS state (red spectrum) are $\Gamma$ = 1.9$\pm$0.4, $kT_s$ = 88$\pm$9 eV, $N$ = 1.1$\pm$0.4 L$_{33}$/d$^2_{10}$, and $\log(A)$ = -3.58$\pm$0.07 ($\chi^2_{red}$=1.06 for 249 dof). A systematic uncertainty of 1\%  represents the instrumental flux calibration uncertainty and  has been applied to all analyzed $Swift$ spectra. 

Analysis of  the $Swift$/XRT data fits (see  Fig.~\ref{three_scal}, pink circles) %points)
 showed that  $\Gamma$  monotonically increases from 1.85 to 2.95, when normalization of the spectral  component (or $\dot M$) increases by a factor of about  15.   We have determined  that energy spectra of SDSS~J0752 in all spectral states can be well modeled using  a product  of the  {\tt tbabs}  and a {\tt BMC} Comptonization component. 

%0987
%---Spectral analysis of SDSS~J0752 in the Comptonization model, %taking into account the FeXXV iron emission line and 
%interstellar absorption, showed a good approximation of the observed spectrum of the source. In this case, SDSS~J0752 was in two spectral states and transitioned between them. An example of changes in the shape and amplitude of the spectrum is presented in Fig. 4. Each of the spectra is characterized by the parameters $\Gamma$, $kT_s$, logA and $N$.

We plotted the dependence of $\Gamma$ on $N_{bmc}$ (Fig.~\ref{three_scal}) and discovered a linear section of monotonic increase in the index with increasing $N$, which is proportional to the accretion rate $\dot M$. It is interesting that the dependence $\Gamma-N_{bmc}$ for high $N_{bmc}$ clearly shows the area of index saturation at the level $\Gamma\sim 3$. This effect is well known as an index saturation, see ST09  (pink line in Fig.~\ref{three_scal}). It was previously tested on a large number of sources with BHs of different masses [stellar-mass BHs \cite{ts09}, \cite{ST10}, \cite{STS14}, \cite{TS23}],  intermediate mass BHs [\cite{TS16}, \cite{tsei16b}] and supermassive BHs [\cite{TS17}, \cite{STV17}, \cite{STU18}, \cite{SCT18}, \cite{TSCO20}, \cite{TSC23}]  and has established itself as a reliable indicator of the BH presence in the source [\cite{ts09}]. %(Titarchuk \& Seifina, 2009). 
In addition to diagnosing the presence of a black hole, this effect makes it possible to estimate the parameters of a black hole object, such as a mass and inclination. 

\subsection{A BH mass estimate of the SDSS~J0752 secondary}
\label{mass_estimate}

We used a previously-developed %our 
technique to estimate the BH mass,  $M_{SDSS}$, 
%previously developed specifically for BH weighing~
\citep{TSC23}. Previously, we did not know a mass of which BH in the SDSS~J0752 quasar can be estimated using our X-ray data for SDSS~J0752. We  used the $\Gamma-N_{bmc}$ correlation to estimate a BH mass (see ST09 for details). This method ultimately ({\it i}) deal with  a pair of BHs for which $\Gamma$ correlates with an {increasing} normalization $N_{bmc}$ (which is proportional to the mass accretion rate $\dot M$ and   a BH mass $M$, cm ST09, Eq.~(7)) and for which the saturation levels $\Gamma_{sat},$ are the same and ({\it ii}) calculates the scaling factor $s_{N}$, which allows us to determine the black hole mass of the target object. It should also be emphasized that to estimate a BH mass  using the following equation for the scale factor, the ratio of the distances to the {\it target} and {\it reference} sources is necessary:
\begin{equation}
 s_N=\frac{N_r}{N_t} =  \frac{m_r}{m_t} \frac{d_t^2}{d_r^2}{f_G},
\label{mass}
\end{equation}
where $N_r$ and $N_t$ are the BMC normalizations of the spectra, $m_t=M_t/M_{\odot}$ and  $m_r=M_r/M_{\odot}$ are the dimensionless  BH masses with respect to solar mass, and $d_t$ and $d_r$  are distances to  the {\it target} and {\it reference} sources, correspondingly. % \LEt{ This is an incomplete sentence, please rephrase.} 
A   geometrical factor, $f_G=\cos i_r/\cos i_t$, where $ i_r$ and $ i_r$ are the disk inclinations for   the {\it reference}  and {\it target} sources, respectively (see ST09, Eq.~(7)).

For  an appropriate  scaling, we have to select X-ray sources (reference sources), which also show the effect of index saturation, and at the same 
$\Gamma$ level as the SDSS~J0752 (target source). For reference sources, a BH's mass, inclination, and distance must be well known. 
We found that OJ~287, M101 ULX--1 and HLX--1 ESO~243 can be used as the reference sources because these sources met all aforementioned  requirements to estimate a BH mass of the target source OJ~287 (see  items (i) and (ii) above). 

\begin{figure}
\centering
\includegraphics[width=9cm]{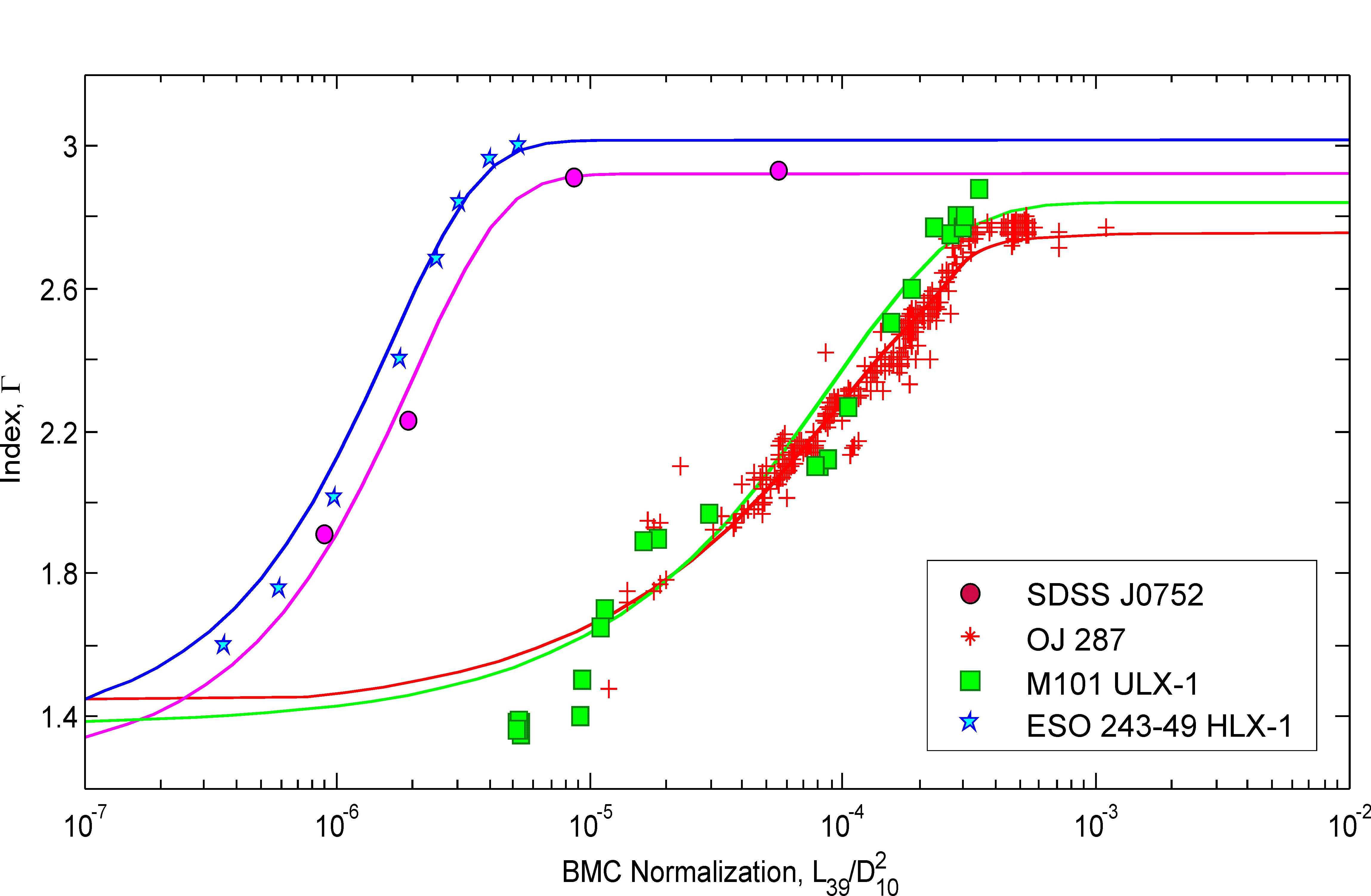} 
  \caption{Scaling of the photon index $\Gamma$ versus  normalization $N_{bmc}$ for SDSS~J0752 (pink  circles %points 
-- target source)  using the correlation for the reference sources OJ~287 (red  crosses), %points), 
ESO~243--49 HLX--1 (blue stars)%points) 
and M101 ULX--1 (green squares).} 
\label{three_scal}
\end{figure}

In Figure \ref{three_scal}  we demonstrate how the photon index $\Gamma$ evolves with normalization $N$ (proportional to the mass accretion rate $\dot M$) in the blazar OJ~287 and in the quasar SDSS~J0752,  where  $N$  is presented in units of $L_{39}/d^2_{10}$ ($L_{39}$ is the source luminosity in units of $10^{39}$ erg/s and $d_{10}$ is the distance to the source in units of 10 kpc).

In Figure \ref{three_scal}, the correlations $\Gamma$ versus $N$ are self-similar for the {\it target} source (SDSS~J0752) and two  M101 ULX--1 and ESO~243--49 HLX--1 are ultra-luminous X-ray (ULX) sources. Moreover, these three sources have almost the same index saturation level $\Gamma$ about 2.8. We estimated a BH mass for SDSS~J0752 using the scaling approach  (see e.g., ST09). In Figure~\ref{three_scal} we illustrate  how the scaling method works shifting  correlations relative to 
% versus 
another.     
From these correlations we could estimate $N_t$, $N_r$ for SDSS~J0752 and   for the reference sources (see Table~\ref{tab:par_scal}). A value of $N_t=5\times10^{-6}$,  $N_r$ in units of $L_{39}/d^2_{10}$  is determined in the beginning of the $\Gamma$-saturation  part [see Fig. \ref{three_scal}, % and \ref{three_scal_1},  
ST07,  ST09, \cite{STS14,TS16,tsei16b,ts09}].
% Seifina et al. (2014); Titarchuk \& Seifina (2016a); Titarchuk \& Seifina (2016b) \& Titarchuk \& Seifina (2009)]. {STS14,TS16,tsei16b,ts09}

To determine the distance to SDSS~J0752 we used the formula (for $z < 1$) 
\begin{equation}
d_{SDSS} = z_{SDSS}~c/H_0\simeq 500~~Mpc, 
\end{equation}
where the redshift $z_{SDSS} = 0.117$ for SDSS~J0752, $H_0 = 70.8 \pm 1.6 km s^{-1} Mpc^{-1}$ is the Hubble constant and $c=3\times 10^5$ km/s is the speed of light. This distance $d_{SDSS}$ agrees with the luminosity distance estimate using Ned Wright's Javascript Cosmology Calculator\footnote{https://www.astro.ucla.edu/~wright/CosmoCalc.html} $d^{NW}_{SDSS}\sim 540$ Mpc \citep{Wright06}. 

A value of   $f_G=\cos {i_r}/\cos{i_t} $  for the {\it target} and {\it reference} sources  can be obtained  using a trial inclination for SDSS~J0752 $i_t=50^{o}$  and for $i_r$   (see Table \ref{tab:par_scal}).
As  a result of the estimated target  mass (SDSS~J0752), $m_t$ we find  that
\begin{equation}
m_t= f_G\frac{m_r}{s_N} \frac{d_t^2}{d_r^2} 
\label{mass_target1}
\end{equation}
where we used  values of $d_t=500$ Mpc. % (see Table \ref{tab:par_scal}).

 Applying   Eq.(\ref{mass_target1}), we can estimate  $m_t$ (see Table \ref{tab:par_scal})  and we  find  that the secondary BH mass  in SDSS~J0752 is about $9\times(1\pm 0.31)\times10^7$ M$_{\odot}$. To obtain this estimate  with appropriate error bars, we need to consider  error bars for $m_r$ and $d_r$ assuming, in the first approximation, errors for  $m_r$ and $d_r$ only.
 We rewrote Eq. (\ref{mass_target1}) as
 \begin{equation}
m_t(1+\Delta m_t/m_t)= f_G\frac{m_r}{s_N} \frac{d_t^2}{d_r^2}(1+\Delta m_r/m_r)(1+ 2 \Delta d_r/ d_r).
\label{mass_target_expand}
\end{equation}

Thus we obtained errors for the $m_t$ determination (see Table \ref{tab:par_scal}, second column for the {\it target} source), such that 
\begin{equation}
\Delta m_t/m_t \sim \Delta m_r/m_r + 2 \Delta d_r/ d_r.
\label{mass_target_errors}
\end{equation} 
As a result, we find that M$_{SDSS} \sim  9\times10^7$ M$_{\odot}$ ($M_{SDSS} = M_t$) assuming $d_{SDSS}= 500$ Mpc to SDSS~J0752. Thus, we obtained a lower limit to the BH mass due the unknown inclination. We present all these results in Table~\ref{tab:par_scal}.

In order to calculate  the  dispersion ${\mathcal D}$ of the arithmetic  mean  $\bar{ m_t}$ for a BH  mass estimate using different reference sources ${\mathcal D}$  (see   Table \ref{tab:par_scal}), one should keep in mind  that 
\begin{equation}
{\mathcal D} (\bar{m_t})= D/n
\label{dispersion_mean}
,\end{equation} 
where $D$ is the dispersion of $m_r$  using  each of the reference sources and $n=3$ is a number of  the reference sources. As a result we determined that  the mean deviation  of the arithmetic mean  
\begin{equation}
\sigma  (\bar{m_t})= \sigma/\sqrt{n}\sim 0.31
\label{sigma_arithm_mean}
\end{equation}
 and finally  we came to the following conclusion (see also Table \ref{tab:par_scal}):
 \begin{equation}
\bar{m_t} \sim9\times(1\pm 0.31)\times10^7 ~~~\rm M_{\odot}. %{solar~masses}. 
\label{arithm_mean}
\end{equation}

It should be noted that in our calculations, we assume the angle between the normal to the secondary disk  and the line of sight to be about  50 degrees. 
However, the actual inclination %this angle 
may be different. %In fact, \cite{Dey21} and \cite{val21} argue that one should see the secondary disk almost face-on, namely, this angle $i_t$ is about zero. Consequently  the mass of the secondary should then be slightly lower, $\bar{m_t} \sim0.8\times10^8$ ~~~\rm {solar~masses}.

%MMM

%-For appropriate scaling, we need to select X-ray sources (reference sources), which also show the effect of index saturation, and at the same $\Gamma$ level as the SDSS~J0752 (target source). For reference sources, the BH's mass, inclination, and distance must be well known. We used OJ~287, M101 ULX--1 and HLX--1 ESO~243 as reference sources.

%-To determine the mass of BH in SDSS using the scaling method, we must know the distance to the SDSS~J0752 source and its inclination. 

\subsection{Secondary disk inclination estimate in SDSS~J0752}
The inclination of SDSS~J0752 is still unknown. However, we can attempt to %we tried to
 estimate it using the scaling method and  the virial BH mass in SDSS~J0752 from optical data \citep{Zhang22} under the assumption that the source of the X-ray variability is a secondary BH. According to \cite{Zhang22}, the virial mass of the secondary BH is about 
$8.8\times 10^7$ M$_{\odot}$. Then using the second scaling law \cite{TSC23}, we find the inclination of secondary BH disk in SDSS~J0752:
\begin{equation}
\label{inclination}
\cos(i_{SDSS}) = \cos(i_r)~ \frac{M_r}{M^{vir}_t} 
\left.\left(
\frac{d_r}{d_t}
\right)
\right. ^{-2}\frac{N_t}{N_r}\sim \cos(80^{\circ}),
\end{equation}
where we used ESO~243--49 HLX--1 as a {\it reference}  source and parameters  $M_r=7.2 \times 10^4$ M$_{\odot}$, $M^{vir}_t=8.8\times 10^7$ M$_{\odot}$, $d_r=95\pm 10$ Mpc, $d_t=500$ Mpc, $N_t=5\times 10^{-6}$ and $N_r=4.2\times10^{-6}$.

Now that we have estimated the inclination of the (mini) disk around the secondary BH based on its virial mass, we can refine our scaling estimate of a BH mass of the secondary. If we previously assumed $f_G=1$, then clarifying the angle $i_t$ leads to $f_G=8$. Since the secondary disk should be seen almost edge-on, namely, this angle $i_t$ tends to 90 degrees, then the mass of the secondary BH component should be slightly larger, $\bar{m_t} \sim7.2\times10^8$ M$_{\odot}$. %{solar~mass}.

\section{DISCUSSION \label{discussion}} 

%-discussion
We estimated the minidisk inclination of the secondary BH by combining scaling of the X-ray spectral properties of  SDSS~J0752 and the virial mass obtained from optical observations (CSS\footnote{ http://nesssi.cacr.caltech.edu/DataRelease/ } and ASAS-SN\footnote{ https://asas-sn.osu.edu/ } V-band light curve) of this quasar~\citep{Zhang22}. A high value of the inclination of the minidisk of the secondary BH was obtained ($\le 80^{o}$). However, it is necessary to emphasize  that the IR and radio data available for SDSS~J0752 are consistent with such a high minidisc inclination. It is known that SDSS~J0752, is a so-called  a blue quasar \citep{Zhang22} is a quasar whose IR radiation is weakly absorbed. As a working hypothesis to explain the status of blue quasars, it was argued that they are observed at a low inclination angle \citep{Klindt19} to the observer and are thus weakly subject to IR absorption. According to this hypothesis, blue quasars are fundamentally different from so-called red quasars, whose radiation is predominantly red at optical wavelengths due to the presence of dust in the line of sight. However, \cite{Klindt19} ruled out orientation as the cause of the differences between red and blue quasars based on an analysis of quasar's radio properties using FIRST data (1.4 GHz/6$\mu$m).

But even in the case of the ``face-on'' orientation of the main disk of the SDSS~J0752 quasar, when we see the galaxy disk  almost from the pole, the high inclination of the secondary BH (mini)disk we obtained is consistent with the fact that the orbit of the secondary BH is oriented perpendicular to the plane of the main disk around the primary BH. This is consistent with our orbital flare scenario (see Fig.~\ref{picture}), in which the orbit of the secondary BH does not coincide with the galactic plane, but makes a significant angle, in our case $\sim 80^{\circ}$. Therefore, we clearly see periodic X-ray and optical flares %outbursts 
during the orbital passage of a secondary BH through the disk around the primary BH (i.e., through the disk of the galaxy).

We obtained a fairly low temperature of the seed photons $kT_s$ from the inner part of the disk around the secondary black hole (80 -- 115 eV). This is consistent with calculations of the X-ray spectrum arising from the innermost part of the source, based on first-principles physical models, taking into account the Comptonization of soft disk photons by hot electrons of the Compton cloud originated from the disk innermost  part %of the source 
and from the converging flow of the black hole. Indeed, in the case of a $10^8 M_{\odot}$ black hole, the peak temperature of the disk is relatively low, $ kT_s<1~{\rm keV}/(M_{BH}/10M_{\odot }) ^{1/4}$ , that is, about 20--100 eV, and its thermal peak is in the UV energy range.
% (for more details about our observations, see %\S 3 and \cite{ss73}).

We proceed with  a scenario of orbital rotation of a binary system in the center of the quasar SDSS~J0752, consisting of two BHs of higher and lower mass (Fig.~\ref{picture}), which explains   its X-ray variability. In this case, a heavier BH %(primary BH) 
is surrounded by a powerful accretion disk, and a lighter BH, %(secondary BH), 
surrounded by %a smaller disk 
the minidisk, rotates around the primary BH in a plane different from the equatorial plane of the accretion disk of the primary BH 
%(according to our calculations, almost perpendicular to it), 
crossing it twice during the orbital period (6.4 years). We also assumed that when a secondary BH passes through the disk around the primary BH, ``tidal disruption'' of nearby parts of the disk occurs.  As a result  a  partial destruction and subsequent replenishment of matter from the accretion minidisk around the secondary BH takes place  (see a similar process in \cite{Chan21}). In this case, a powerful transition minidisk is  developed around the secondary BH with subsequent accretion of material from the transition minidisk onto the secondary BH. This provides an increase in luminosity % in the form of a burst 
in the X-ray/optical/radio bands.

To complete our analysis, we list other possible reasons leading to the variability of quasar emission. It should be also  taken into account that this variability of quasars may be associated with changes in emission of 
%the glow of 
their jets.   This source is a weak radio emitter and  is not classified as "radio loud", By conventional definition of radio loudness it  is $R=L_{radio}/L_{bolometric}\sim 10$.
 For this object its $R\sim0.3$.  According to a recent structural analysis 
 of 447 of the 
 %brightest 
 radio
 %loud 
 sources 
 in the northern sky at 15 GHz \citep{Lister21}, variations in the position angle of the jets 
 %are quite often, 
 with an average amplitude of 10--50$^{\circ}$ on a timescale of about ten years. For some sources, variations reach 200$^{\circ}$. There are several scenarios that could explain this behavior, see for example, orbital motion in a binary black hole \citep{Begelman80}. The Lense--Thirring effect \citep{Lense+Thirring18,Thirring18} in a binary system with a rotating black hole is  widely discussed. In this case, fluctuations in the internal orientation of the jet are caused by the precession of the accretion disk, the rotation axis of which is shifted relative to the rotation axis of the black hole \citep{Caproni04}. For example, based on observations made over 22 years in 170 separate epochs, a study of the structure of the jet in the radio galaxy M87 showed a periodic change in the position angle of the jet with a peak-to-peak amplitude
of $\sim$10$^{\circ}$ and a period of T$\sim$11 years on scales $\sim$600--2500 $r_g$ \citep{Cui23}. The periodicity of the light curve in the optical range and variations in the position angle of the jet in the quasar 3C~120 suggest the existence of a precession of the jet caused by the Lense--Thirring effect with a period of 12.3 years \citep{Caproni+Abraham04}. The same is also  shown for the jet in M81$\ast$, the precession period of which is estimated at $\sim$7 years \citep{Fellenberg23}. A detailed analysis of the orientation oscillations of other jets also indicates their periodicity and possible connection with precession, for example,  
PKS~2131--021 \citep{ONeill22} ($T\sim$ 22 years), 
PG~1553+113 \citep{Lico20}, 4C~38.41 \citep{Algaba19} ($T = 23 \pm 5$ years), 
3C~279 \citep{Abraham+Carrara98} ($T\sim$ 22 years), 
3C~273 \citep{Abraham+Romero99} ($T\sim$ 16 years ) and 
OJ~287 \citep{Britzen18} ($T\sim$ 12 years). This list may be supplemented by SDSS~J0752, but this requires an  additional research.

One can argue 
%that we did not realize 
that the emission from this quasar object comes from a jet. However, we do not see any serious arguments for this statement. A particularly important point regards the fitting of $Swift$ X-ray spectra.  One could think that  our spectral  models contain too many components and thus,  that they could not be fitted to low-resolution $Swift$ data,  since there would then be many more free parameters than actual independent data bins.  This is not the case because our continuum spectral model is  an XSPEC model consisting of the BMC component. The spectral model parameters are  equivalent hydrogen absorption column density $N_H$; the photon index $\Gamma$; $\log(A)$, which is related to the Comptonized factor $f$;  and the color temperature and normalization of the seed photon blackbody component $kT_s$ and $N$, respectively. 

Since the more massive BH is not active in the X-rays, then the active culprit for the X-ray flares, for example, in M87 is precisely the secondary (less massive) BH.  It is this activity that underlies the X-ray scaling and timing analysis methods, and their results relate to the smaller black hole of this binary system in M87  and  now it is not surprising that a BH mass according to X-ray estimates by~\cite{TSCO20} turned out to be two orders of magnitude less than results of the EHT method~\citep{Akiyama19} and classical gas and stellar dynamic approach~\citep{Gebhardt11,Walsh13}.
%\subsection{Comparison with other results  in the literature}

%000000
\section{CONCLUSIONS}

The optical periodic variability of the SDSS~J0752 quasar has raised several questions regarding whether the source contains one or two BHs. It was very important to identify the X-ray characteristics indicating the duality of  this quasar. In this paper, we demonstrated that an X-ray source coupled to an optical one exhibits flux variability in the energy range of 0.3--10 keV by a factor of 15. At the same time, the X-ray spectra of SDSS~J0752 undergo transitions from the LHS state to the IS and then to the HSS (see Fig.~\ref{3_spectra}) based on data from the XRT telescope on board the {\it Swift} observatory. We fitted the  SDSS~J0752 energy spectrum  to a Comptonization model and described the evolution of the spectrum parameters. The temperature of the seed disk photons turned out to be very low ($kT_s=0.08-0.15$ keV), the degree of disk  illumination  varied in a wide range (illumination factor $f=0.1-1$) and the disk luminosity in the soft X-ray range $L_{0.3-10keV}$ showed a monotonic increase during the state evolution -- the LHS$\to$IS$\to$HSS --  with a factor of ten and correlated with the optical flash of the SDSS~J0752 ($N_{bmc} = (0.1-8)\times L_{34}/d^2_{10}$). In addition, we discovered saturation of the photon index with  the mass accretion rate in SDSS~J0752, which allowed us to scale the secondary black hole in this object using the scaling method.
% (9$\times$10$^7$ M$_{\odot}$). 
%For scaling, we used the massive BH sources OJ~287, M101 ULX--1 and HLX--1 ESO--243 as reference sources.
%This is our main result.
 We thus determine mass of 9$\times$10$^7$ M$_{\odot}$ for the secondary BH.
 We have shown that a scenario in which two supermassive black holes at the center of the quasar SDSS~J0752 form an orbital pair, and the less massive secondary BH periodically crosses/pierces the disk around the more massive primary BH, can be applied to SDSS~J0752. Thus, we associated an increase of the X-ray flux in SDSS~J0752 with  intersections of the disk of the primary BH with the secondary BH. In addition, we estimated the orbital inclination of the secondary BH, $i=80^{\circ}$, based on a combination of our scaling approach  and virial BH mass estimates by \cite{Zhang22}. % (Fig. 1).

\section*{ACKNOWLEDGMENTS}
%\section*{Acknowledgments}
We thank the anonymous referee for the careful reading of the manuscript and for providing valuable comments.  %In particular, 
The Authors are very  happy to get a careful reading  and editing our  manuscript by Chris Shrader.   We thank the anonymous referes for the careful reading of the manuscript and for providing valuable comments.  We acknowledge support from UK $Swift$ Science Data Centre at the University of Leicester for supplied data. 
 This paper has made use of the data from the CSS projects http://nesssi.cacr.caltech.edu/DataRelease/.

%{\bf Author Contributions}

%Conceptualization, methodology, L. T. ; data analysis, E. S.; investigation and writing—original draft preparation, L. T. and E. S.

%{\bf Funding}

%This research received no external funding.

\section*{CONFLICTS OF INTEREST}
The authors declare no conflict of interest.

%\section*{Conflict of Interest Statement}
%All financial, commercial or other relationships that might be perceived by the academic community as representing a potential conflict of interest must be disclosed. If no such relationship exists, authors will be asked to confirm the following statement: 

%The authors declare that the research was conducted in the absence of any commercial or financial relationships that could be construed as a potential conflict of interest.

%The authors of this work declare that they have no conflicts of interest.

%\section*{Author Contributions}

%The Author Contributions section is mandatory for all articles, including articles by sole authors. If an appropriate statement is not provided on submission, a standard one will be inserted during the production process. The Author Contributions statement must describe the contributions of individual authors referred to by their initials and, in doing so, all authors agree to be accountable for the content of the work. Please see  \href{https://www.frontiersin.org/about/policies-and-publication-ethics#AuthorshipAuthorResponsibilities}{here} for full authorship criteria.

\section*{Funding}

%This work was supported by ongoing institutional funding. 
No additional grants to carry out or direct this particular research were obtained.
%Details of all funding sources should be provided, including grant numbers if applicable. Please ensure to add all necessary funding information, as after publication this is no longer possible.

%\section*{Acknowledgments}
%This is a short text to acknowledge the contributions of specific colleagues, institutions, or agencies that aided the efforts of the authors.

\section*{Data Availability Statement}

%The data for Insight-HXMT underlying this article is available format at Insght-HXMT website (http://archive.hxmt.cn/proposal).
The {\it Swift} and {\it CSS} data underlying this paper are publicly available through the GSFC public archive (http://heasarc.gsfc.nasa.gov) and the CSS project (http://nesssi.cacr.caltech.edu/DataRelease/).

\newpage

%%%%%%%%%%%%%%%%%%%%%%%%%%%%%%%%%%%%%%%%%
%
% TABLE  0 - for basic parameters of SDSS J0752          ....................1
%
%%%%%%%%%%%%%%%%%%%%%%%%%%%%%%%%%%%%%%%%%
\begin{table}%[htb]
%% use tabular font for a smaller size font
\centering
%\tabularfont
\caption{Basic parameters of SDSS~J0752.}
\label{tab:parameters_sdss}
 \begin{tabular}{llrcl}
 \hline\hline                        % inserts double horizontal lines
  Parameter               &     Value     &   Reference      \\
      \hline
Source class                                     & QSO                                & \cite{Zhang22}             \\
%Mass of an X-ray star, M$_{\odot}$ & $\sim$1.4~\cite{Giles02}              &1.19 $\pm$ 0.02~\cite{Friend90},   0.81$ \pm$ 0.19~\cite{Bitner07}\\
%an X-ray star, %M$_1$,  
%M$_{\odot}$                                                                &                            & 0.81$ \pm$ 0.19$^{(c)}$  \\
%Spectral type  of secondary    & ...                         & K5/5 \\
% of secondary                       &                          &  \\
Virial mass of a primary BH, M$_{\odot}$ & $\sim$1.04$\times 10^9$     &\cite{Zhang22} \\
Virial mass of a secondary BH, M$_{\odot}$ & $\sim$8.8$\times 10^7$   &\cite{Zhang22} \\
Orbital inclination,  $i$,  deg               & $\sim$ 80                             &this work\\
Orbital period,  $P$,  yr                      & 6.4                                 &\cite{Zhang22}  \\
Distance $D$,  Mpc                            & $\sim$ 500                      & this work\\
Red shift, $z$                                      & 0.117                             &\cite{Paris18} \\
 \hline                                             %inserts single line
\end{tabular}
%%use \tablenotes{footnote} to get the table foot note
%\tablenotes{
%\\$^{(a)}$ \cite{Giles02}; % Giles, A. B., Hill, K. M., Strohmayer, T. E., & Cummings, N. 2002, ApJ, 568, 279
%$^{(b)}$ \cite{Friend90}; 
%$^{(c)}$ \cite{Bitner07}; %Bitner, Robinson, and Behr (2007).
%$^{(d)}$ \cite{Sanna13,Casares06,Frank87}; %Frank et al. 1987; Casares et al. 2006; Sanna et al. 2013;
%$^{(e)}$ \cite{Ritter_Kolb03}; %Tominaga et al. 2020
%$^{(f)}$ \cite{Galloway20}; 
%$^{(g)}$ \cite{Harrison04};  
%$^{(h)}$ \cite{smale_mukai88} and \cite{van_Paradijs90}.
%}%%9/11
\end{table}

%%%% 0000000 TABLES SDSS    ......................................................1

%%%%%%%%%%%%%%%%%%%%%%%%%%%%%%%%%%%%%%%%%%%%%%%%%%%%
%
%   TABLE   1  Swift data
%
%%%%%%%%%%%%%%%%%%%%%%%%%%%%%%%%%%%%%%%%%%%%%%%%%%%%
%%%%Lena
\begin{table}
 \caption{List of the $Swift$ observations of SDSS~J0752 used in our analysis.}              % title of Table
 \label{tab:list_Swift}      % is used to refer this table in the text
% \centering                                      % used for centering table
 \begin{tabular}{l l l c r c}          % centered columns (4 columns)
 \hline\hline                        % inserts double horizontal lines
  Obs. ID& Start time (UT)  && Exposure time, ks  &Start time (MJD)  \\    % table heading
 \hline                                   % inserts single horizontal line
00038041001   & 2008 October 31 07:06:59  && 3.6    & 54770.297 \\%&\\
00038041002 & 2009 February 24 04:55:17  && 1.3    & 54886.205  \\%  & \\
00038041003 & 2009 September 12 19:12:59 && 1.5 & 55086.800 \\%& \\
00038041004 & 2010 May 30 00:07:16          && 2.7     & 55346.005 \\%& \\
00039551001 & 2010 May 30 09:44:39          && 5.0  & 55346.406 \\%&\\
 \hline                                             %inserts single line
 \end{tabular}
 \end{table}

%For PLOT
%-%OLD-------------------------------------------------------------------------------------------------------------------------------------------------------------------
%N                MJD    alpha   err_  T_s,  err_  log(A)  err     norm    err_   N_H err_ xi2(dof)  cnt/s err  expo 
%                               alpha  keV   T_s           logA    BMC     norn      N_H                      (sec)
%                                                                  x10^{-2}!!!   x10^{22}!!!   %-----------------------------------------------------------------------------------------------------------------------------------------------------------------
%38041001    54770.297   2.00  0.63  0.025  0.01   2.00   50.0    8.64   0.3    4.4  0.37  0.47 926  0.055  0.002  1468
%38041002    54886.205  2.00  0.63  0.059  0.01   2.00   50.0    5.07   0.3    3.94 0.65  0.23 965  0.055  0.002  1468
%38041004    55346.005  1.01  0.63  0.071  0.01  -0.87   0.06    6.11   0.3    1.12 0.53  0.43 964  0.055  0.002  1468
%38041003    55086.800  1.01  0.63  0.071  0.01  -0.87   0.06    6.11   0.3    1.12 0.53  0.43 964  0.055  0.002  1468 %% bad
%39041001    55346.406  1.55  0.10  0.029  0.04   2.00   50.0    3.29   0.3    6.84 0.29  0.18 964  0.055  0.002  1468
%--------------------------------------------------------------------------------------------------------------------------------------------------------------------

%%%%%%%%%%%%%%%%%%%%%%%%%%%%%%%%%%%%%%%%%%%%%%%%%%%%%%%%%%%%%%%
%
% TABLE 2 - BH MASS DETERMINATION
%
%%%%%%%%%%%%%%%%%%%%%%%%%%%%%%%%%%%%%%%%%%%%%%%%%%%%%%%%%%%%%%%
\begin{table}
 \caption{BH masses and distances.}
 \label{tab:par_scal}
 \centering 
 \begin{tabular}{lllrrll}
 %\begin{tabular}{llllllll}
 \hline\hline                        % inserts double horizontal lines
%Reference sources   & $m_r$ (M$_{\odot})$ & $i_r^{(a)}$ (deg) & $N_r$ ($L_{39}/d^2_{10}$) & $d_r^{(b)}$ (Mpc) \\
Reference    & $m_r$,         & $i_r^{(a)}$,  & $N_r$,                     & $d_r^{(b)}$,  \\
sources       & M$_{\odot}$ &  deg             &  $L_{39}/d^2_{10}$ &  Mpc \\
      \hline%\hline
%XTE~J1550--564$^{(1)}$ &   10.7$\pm1.5$ &  72  & 1. &  3.3 $\pm0.5$ \\       %\cite{st09}  ST09
%H ~1743--322$^{(2)}$  &  13.3$\pm3.2$ & 70 & 0.19 &  $9.1\pm1.5$\\   
%4U~1630--47$^{(3)}$ & 10$\pm 0.1$& 70& 0.12 & $10\pm 1$\\
%HLX--1 ESO 243--49$^{(1)}$ & $(7.2\pm 0.7)\times10^4$ & 75 & $4.2\times 10^{-6}$ & $ 95\pm 10$\\
HLX--1                                &  &  & \\
~~ESO 243--49$^{(1)}$ & $(7.2\pm 0.7)\times 10^4$ & 75 & $4.2\times 10^{-6}$ & $ 95\pm 10$\\
M101 ULX--1$^{(2)}$ & $(3.7 \pm 0.6)\times 10^4$ & $18$ & $3\times 10^{-4}$ & $6.9\pm 0.7$\\
OJ 287$^{(3)}$ & $(1.25 \pm 0.5)\times 10^8$ & $50$ & $2.4\times 10^{-4}$ & $1037\pm 10$\\
%GRS 1915+105 $^{(6)}$ & $12.4\pm 2$ & $70$ & $0.2$ & $8.6\pm 2 $\\
 \hline\hline                        % inserts double horizontal lines
%Target source   & $m_{t}$ (M$_{\odot}$) & $i_t^{(a)}$ (deg) & $d_t^{(b)}$ (Mpc) &    \\
Target source   & $m_{t}$,      & $i_t^{(a)}$,  & $d_t^{(b)}$, &    \\
                      & M$_{\odot}$ &  deg             & Mpc &    \\
      \hline
%OJ 287 & $\sim1.25\times(1\pm 0.45)\times10^8$ &   50  &  1.037 Gpc     &   that  using  XTE~J1550-564- as a reference source \\
% OJ 287  & $\sim1.25\times(1\pm 0.45)\times10^8$ &  50 &  1.037 Gpc    &    that  using  H ~1743--322  as a reference source \\ 
%OJ 287  & $\sim1.25\times(1\pm 0.45)\times10^8$ &  50 &  1.037 Gpc    &    that  using  4U~1630--47  as a reference source\\
SDSS~J0752  & $\sim9\times(1\pm 0.45)\times10^7$ &  50 &  500    &    that  using  ESO 243-49  as a ref. source\\
SDSS~J0752  & $\sim9\times(1\pm 0.45)\times10^7$ &  50 &  500    &    that  using  M101 ULX-1  as a ref. source\\
%OJ 287  & $\sim1.25\times(1\pm 0.45)\times10^8$ &  50 &  1.037 Gpc    &    that  using  GRS 1915+105   as a reference source\\
SDSS~J0752  & Final  estimate                                      &  50 &  500   &   as a standard deviation for a mean: \\% $0.45/6^{1/2}=0.18$\\
           &  $\sim9\times(1\pm 0.31)\times10^7$ &      &      &   $0.45/3^{1/2}=0.31$\\
\hline
%\\%, $\ge$ 15$^e$\\%$\sim$ 7--21 \\
% \hline                                             %inserts single line
 \end{tabular}
\\
(a)  System inclination in the literature and  
(b) source distance found in the literature. %(c) scaling value found by ST09; 
%(c) scaled value found by $\Gamma-\nu_{L}$ correlation and (d) scaled value found by $\Gamma-N_{bmc}$ correlation of the present paper.
%(1) ST09;  \cite{Orosz2002,SanchezFernandez99,Sobczak99};
%(2) ST09;
%(3) \cite{STS14}; % Seifina et al. (2014);
(1) \cite{tsei16b}; %  Titarchuk \& Seifina (2016a);
(2)  \cite{TS16}; and %Titarchuk \& Seifina (2016b);
(3) \cite{TSC23}.
%(6) \cite{ts09}. %Titarchuk \& Seifina (2009).
%\\
 \end{table}

%%%%1111111111111 FIGURES

%%%%%%%%%%%%%%%%%%%%%%%%%%%%%%%%%%%%%%%%%%%%%%%
%
% FIGURE 1
%
%%%%%%%%%%%%%%%%%%%%%%%%%%%%%%%%%%%%%%%%%%%%%%%%%%%%

%%%%%%%%%%%%%%%%%%%%%%%%%%%%%%%%%%%%%%%%%%%%%%%
%
% FIGURE 2
%
%%%%%%%%%%%%%%%%%%%%%%%%%%%%%%%%%%%%%%%%%%%%%%%%%%%%

%%%%%%%%%%%%%%%%%%%%%%%%%%%%%%%%%%%%%%%%%%%%%%%
%
% FIGURE 3
%
%%%%%%%%%%%%%%%%%%%%%%%%%%%%%%%%%%%%%%%%%%%%%%%%%%%%

%%%%%%%%%%%%%%%%%%%%%%%%%%%%%%%%%%%%%%%%%%%%%%%
%
% FIGURE 3+
%
%%%%%%%%%%%%%%%%%%%%%%%%%%%%%%%%%%%%%%%%%%%%%%%%%%%%

%%%%%%%%%%%%%%%%%%%%%%%%%%%%%%%%%%%%%%%%%%%%% 
%
%  FIGURE 4 spectra
%
%%%%%%%%%%%%%%%%%%%%%%%%%%%%%%%%%%%%%%%%%%%%%          

%%%%%%%%%%%%%%%%%%%% 
%
%  FIgure 5 - GAMMA - NORM
%
%%%%%%%%%%%%%%%%%%%% 

%%%%%%%%%%%%%%%%%%%%%%%%%%%%%%%%%
% 
%  FIgure 6 - SCALING 1550
%
%%%%%%%%%%%%%%%%%%%%%%%%%%%%%%%%%

\end{document}